\documentstyle[times, emulateapj]{article}

\def\be{\begin{equation}}
\def\ee{\end{equation}}
\def\bea{\begin{eqnarray}}
\def\eea{\end{eqnarray}}
\def\ri{r_{I}}
\def\ro{r_{O}}
\def\aproxgt{\mathrel{%
      \rlap{\raise 0.511ex \hbox{$>$}}{\lower 0.511ex \hbox{$\sim$}}}}
\def\aproxlt{\mathrel{%
      \rlap{\raise 0.511ex \hbox{$<$}}{\lower 0.511ex \hbox{$\sim$}}}}

\received{November 4, 1996}
\accepted{December 12, 1996}

\slugcomment{Submitted to The Astrophysical Journal, Letters}

\lefthead{Nowak, Wagoner, Begelman, \& Lehr}
\righthead{The 67 Hz Feature in Black Hole Candidate GRS 1915+105 $\ldots$}

\begin{document}

\title{The 67 Hz Feature in the Black Hole Candidate GRS 1915+105 as a 
Possible \break ``Diskoseismic" Mode}

\author{Michael A. Nowak\altaffilmark{1}, Robert
V. Wagoner\altaffilmark{2}, Mitchell C. Begelman\altaffilmark{1,3}, Dana
E. Lehr\altaffilmark{2} }

\altaffiltext{1}{JILA, University of Colorado, Campus Box 440, Boulder, CO
~80309-0440}

\altaffiltext{2}{Dept. of Physics, Stanford University, Stanford, CA
~94305-4060}

\altaffiltext{3}{also Dept. of APAS, University of Colorado, Boulder, CO 
~80309}

\affil{\sl Received November 4, 1996; accepted December 12, 1996}

\begin{abstract}

The {\it Rossi X-ray Timing Explorer} ({\it RXTE}) has made feasible for
the first time the search for high-frequency ($\aproxgt 100$ Hz) periodic
features in black hole candidate (BHC) systems.  Such a feature, with a 67
Hz frequency, recently has been discovered in the BHC GRS 1915+105 (Morgan,
Remillard, \& Greiner).  This feature is weak (rms variability $\sim 0.3\%
-1.6\%$), stable in frequency (to within $\sim 2$ Hz) despite appreciable
luminosity fluctuations, and narrow (quality factor $Q \sim 20$).  Several
of these properties are what one expects for a ``diskoseismic" $g-$mode in
an accretion disk about a $10.6~ M_\odot$ (nonrotating)$ - 36.3~ M_\odot$
(maximally rotating) black hole (if we are observing the fundamental mode
frequency).  We explore this possibility by considering the expected
luminosity modulation, as well as possible excitation and growth
mechanisms---including turbulent excitation, damping, and ``negative"
radiation damping. We conclude that a diskoseismic interpretation of the
observations is viable.

\end{abstract}

\keywords{black hole physics --- X-rays: Stars}

\section{Introduction}

In a series of previous papers (Nowak \& Wagoner 1992; Nowak \& Wagoner
1993; Perez et al. 1997; hereafter NW92, NW93, P97, respectively), we have
discussed a class of modes in thin, Keplerian accretion disks that only
exist in the presence of strong-field gravity (not in Newtonian gravity),
some of whose eigenfrequencies depend primarily upon the mass and angular
momentum of the black hole.  The modes are trapped by general-relativistic
modification of the radial ($\kappa$) and vertical ($\Omega_\perp$)
epicyclic frequencies of free particle, circular orbit perturbations (P97,
and references therein).

These modes of oscillation are perturbations of the disk which are
proportional to $\exp[i(\sigma t + m \phi)]$. With disk angular velocity
$\Omega(r)$, their corotating frequency is $\omega(r) = \sigma + m \Omega$.
Three classes of modes ($p-$, $g-$, and $c-$modes) have been identified
(P97, and references therein).  However, here we shall be concerned solely
with radial ($m=0$) $g-$modes, which are the most relevant
observationally\footnote{These modes cover the largest area of the disk in
a region where the disk temperature is maximum (away from the uncertain
conditions at the disk edge).}. These $g-$modes are trapped where $\omega^2
< \kappa^2$, in the region where $\kappa$ achieves its maximum value (at
$r=8~ GM/c^2$ for a nonrotating black hole) because, unlike in Newtonian
gravity, $\kappa(r)$ rolls over at small $r$ and vanishes at the inner disk
edge, thereby creating a resonant cavity.  The modes with the fewest radial
nodes have relatively large radial extents $\Delta r \approx GM/c^2$.

The frequencies ($f=-\sigma/2\pi$) of the
radial $g$--modes are 
\bea 
f~=  &714& \left ( {{M_\odot}\over{M}} \right ) ~ F(a) ~ [1-\epsilon_{nj}] 
~~{\rm Hz} ~ , 
\nonumber \\
&\epsilon_{nj}&  \sim~ \left ( ~ {{ n+1/2 }\over { j+\delta }}~ \right ) 
      \left ( {{h} \over {r}} \right ) ~.  
\label{frequency}
\eea (cf. P97).  
Here $F(a)$ is a known function of the dimensionless black
hole angular momentum parameter $a=cJ/GM^2$, ranging from $F(0) = 1$ to
$F(0.998) = 3.443$.  The properties of the disk enter only through the
small correction term involving the disk thickness $2h$, and the radial
($n$) and vertical ($j$) mode numbers, with $\delta \sim 1$ (as derived
from the WKB solutions of P97).  For thin disk models where $h/r \sim 0.1
~L/L_{Edd}$ (in the mode trapping region), $\epsilon_{nj}$ is typically on
the order of a percent for the lowest modes ($n \sim j \sim 1$). Therefore,
the mode frequency is relatively independent of disk luminosity.
(The mode width, however, scales as $\Delta r \propto \sqrt{c_s}
\propto \sqrt{h}$, where $c_s$ is the sound speed in the disk.)  The radial
$g-$mode is thus a good candidate for explaining the recently discovered
high-frequency features in the black hole candidate GRS 1915+105.

The observed frequency of this feature is 67 Hz, with a full width half
maximum of $3-4$ Hz.  (Morgan, Remillard, \& Greiner 1996, 1997).  This
implies an effective $Q = f/ \Delta f \sim 20$.  Despite factors of
$\sim 2$ luminosity variations in the source, the frequency remained
constant to within $1-2$ Hz. The root mean square (rms) variability of the
67 Hz feature varied from $0.3\% - 1.6\%$ of the {\it total} observed X-ray
luminosity, with the lower limit essentially being the Poisson noise limit.
At its strongest, the rms variability $\sim 1\%$ below 6 keV, and the rms
variability $\sim 6\%$ in the $12-25$ keV band, indicating that the
spectrum of the feature is harder than the integrated source spectrum. (The
spectrum of the source is fit by a power-law exponentially cut-off at $\sim
5$ keV.)

Below we discuss to what extent $g-$modes can explain these observations.
In \S 2 we further review the properties of these modes, and make simple
estimates of their luminosity modulation.  In \S 3 we discuss
application to GRS 1915+105.  In \S 4 we discuss the role of turbulence and
radiation as possible damping, growth, or excitation mechanisms.
In \S 5 we present our conclusions.

\vfill\eject
\section{Luminosity Modulation of Modes}

For both the psuedo-Newtonian (NW92, NW93) and fully relativistic (P97)
diskoseismology calculations, the equations are formulated in terms of the
potential $V \equiv \delta P/\rho$, where $\delta P$ is the Eulerian
variation of the pressure and $\rho$ is the unperturbed density. For 
modes with an azimuthal and time dependence $\propto \exp[i(m\phi + \sigma
t)]$, the potential $V(r,\eta, \phi, t)$ can be separated into a radial,
$V_r(r)$, and a vertical, $V_\eta(\eta)$, component, where $\eta \equiv
z/h(r)$ is the dimensionless vertical coordinate.  The (small) fluid
displacement vector, $\vec \xi$, can be related to the
Eulerian potential $V(r,\eta, \phi, t)$ via the equations 
\bea
~~\xi_r(r,\phi,z,t) ~\approx~ (\omega^2 - \kappa^2)^{-1} ~\exp[i ( m \phi 
    + \sigma t )] ~ V_\eta {{\partial V_r}\over{\partial r}} \quad
     \nonumber \\
~\xi_z (r,\phi,z,t) ~\approx~ (\omega^2 - N_z^2)^{-1} \exp[i ( m \phi 
     + \sigma t) ] 
    ~ V_r {{\partial V_\eta}\over{\partial z}} ~~, 
\label{lagrange}
\eea where $N_z(\eta)$ is the vertical buoyancy frequency, which 
we shall set to zero (cf. P97 for a discussion of the role
of non-vanishing bouyancy frequency).  We shall usually use $\Delta$
to denote a {\it Lagrangian} variation of a given quantity, where we follow the
definitions of NW92, NW93, and P97.  (For $m=0$, $g-$modes have  
$\omega^2 = \sigma^2 \aproxlt \kappa^2$.)

The radial component of the fluid displacement approximately
satisfies the WKB relation
\be
\omega^2 c_s^2(r, 0) {{d^2 W}\over{d r^2}} ~+~
     (\omega^2 - \kappa^2) (\omega^2 - \Upsilon \Omega_\perp^2) ~ W ~=~ 
     0 ~~,
\label{dispersion}
\ee where $\Upsilon(r)$ is a slowly varying separation function (akin to a
separation constant in the WKB limit),
$c_s(r,0)$ is the speed of sound at the disk midplane, 
$W \equiv (\kappa^2-w^2)^{-1} d V_r/d r$, and 
$\Omega_\perp = \Omega$ for a nonrotating black hole.  We have taken the 
perturbations to be adiabatic, which means that the Lagrangian perturbation
of the pressure and density are given by 
${{\Delta P}/{P}} = \gamma {{\Delta \rho} / {\rho}} \approx -\gamma \vec 
\nabla \cdot \vec \xi $.  Utilizing the radial WKB approximation employed
in equation (\ref{dispersion}), the (approximate) components of the divergence 
become 
\bea 
{{\partial \xi_r}\over{\partial r}} &\approx& \big [ {{(\Upsilon \Omega^2 
- \omega^2)} / {h^2 \Omega^2 \omega^2}} \big ]~ \exp(i \sigma t) 
~ V_r(r) V_\eta(\eta) 
\nonumber \\
{{\partial \xi_z}\over{\partial z}} &\approx& (h \omega)^{-2} ~~
     \exp(i \sigma t) ~ V_r(r) {{\partial^2 V_\eta(\eta)} \over {\partial
     \eta^2}} ~. 
\label{diskoII} \eea
(We have also taken the unperturbed vertical barotropic structure
to be in hydrostatic equilibrium.)

For most simple $\alpha$-models, the energy generation rate per unit volume
is approximately $\alpha P(r,z) \Omega(r)$.  The modes not only
perturb the pressure (we ignore possible perturbations to $\alpha$, and one
can show that perturbations to $\Omega$ are negligible), but they also
perturb the locations of the disk boundaries.  The variation of the
luminosity is therefore 
\bea 
\delta L &\sim& 2 \pi \Bigg [
\int_{\ri+\xi_r(\ri)}^{\ro+\xi_r(\ro)} r' dr'
\int_{-z_0+\xi_z(-z_0)}^{z_0+\xi_z(z_0)} \alpha \Omega P'(r',z') dz'
\nonumber \\ &\mbox{}& - \int_{\ri}^{\ro} r dr ~ \int_{-z_0}^{z_0} \alpha
\Omega P(r,z) ~dz \Bigg ] ~ ,
\label{deltaL}
\eea
where $P'(r',z') \equiv P(r,z) + \Delta P(r,z)$, $r' \equiv r+
\xi_r(r)$, $z'=z+\xi_z(z)$, and $\ri$, $\ro$, $z_0$ are the disk boundaries
($\ro$ can be taken to go to $\infty$ without loss of generality).
Transforming variables, the change in luminosity can 
be written as
\goodbreak
\bea
\delta L &\sim&  2 \pi \int_{\ri}^{\ro} dr~ \alpha \Omega(r) r 
     \int_{-z_0}^{z_0} dz~ P(r,z) \qquad
\nonumber \\
     &\times& \bigg \{~ (1-\gamma) \vec \nabla \cdot \vec \xi
 	~+~ \left [ {{\partial \xi_r}\over{\partial r}} 
     {{\partial \xi_z}\over{\partial z}} - \gamma \left ( \vec \nabla \cdot
     \vec \xi \right )^2 \right ] 
\nonumber \\
     &\mbox{}& ~-~ \gamma {{\partial \xi_r}\over{\partial r}} 
     {{\partial \xi_z}\over{\partial z}} \vec \nabla \cdot \vec \xi  ~\bigg \}
     ~~, 
\label{eulerL}
\eea
where we have employed the above expression for the Lagrangian 
pressure change, and employed the WKB approximation in 
{\it each} harmonic term above.

The luminosity fluctuation in general will have contributions
from the fundamental frequency, as well as the first two harmonics.
We have calculated the
luminosity modulation for modes in a disk with radial sound speed profile
equal to that of a radiation-pressure dominated Shakura-Sunyaev disk
with $L/L_{Edd} = 0.3$.  The exact {\it radial} profile does not greatly
effect the estimates, as we can replace the quantity $\smallint P(r,0)
\Omega(r) dz$ with $F(r)$, the disk energy flux, which is known from energy
conservation.  As the mode width $\Delta r \propto
\sqrt{h}$, the magnitude of $h$ {\it does} affect the
luminosity estimates.  If we define the {\it maximum}
magnitude (achieved at a radius, $r_m$) of the vertical
displacement vector $\xi_z(r_m,z,t)$ to be $\equiv {\cal H} h(r_m)$, then
for the case of the mode with one radial node in $V_r$ and two vertical
nodes in $V_\eta$ ($\partial \xi_z/\partial \eta$ even about the
midplane)\footnote{The first ``even" mode about the midplane is not trapped
in the region of the epicyclic frequency maximum. The first ``odd" mode
leads to a vanishing vertical integral in the above estimate of luminosity
modulation.}  we have 
\bea 
\left ( {{\delta L}\over{L}} \right ) &\approx&
0.6\% ~ {\cal H} ~ \sin ( \sigma t + \theta ) ~-~ 9.8\% ~ {\cal H}^2 ~
\sin^2 ( \sigma t + \theta ) \nonumber \\ &\mbox{}& +~~ 0.2\% ~ {\cal H}^3
~ \sin^3 ( \sigma t + \theta ) ~.
\label{deltaL/L}
\eea
(Here we have used the psuedo-Newtonian 
approximation of NW92, NW93 for our calculations.) This is the
modulation of the {\it bolometric} luminosity.  As the mode exists in the
inner, hotter regions of the accretion disk, the
modulation in restricted high energy bandpasses will be greater.
However, we have not properly accounted for radiative transfer effects.
Dispersion in photon diffusion times over the extent of the mode will
decrease the modulated luminosity for large optical depths.

\section{Application to GRS 1915+105}

Although the above luminosity modulation is not large, ${\cal H} \sim
0.6-1.6$ produces, at the fundamental frequency, rms variability $\sim
0.3-1.6\%$ [ $=(0.6 \% {\cal H} + 0.2 \% {\cal H}^3)/\sqrt{2}$ ], as is
seen for the feature in GRS 1915+105.  The first harmonic can produce this
observed rms variability for ${\cal H} \sim 0.3-0.7$.  (The rms variability
is the coefficient of the $\sin^2$ term divided by $\sqrt 8$.)  The
modulation seen in GRS 1915+105 increases to 6\% if the energy band is
restricted to the $12-25$ keV range.  This qualitatively agrees with what
we expect for our modes.  If we compare the bolometric luminosity variation
of the mode to the bolometric luminosity from the region $r \approx 6 - 20
~GM/c^2$, the relative fractional variation increases to $\sim 6\%$.

For the {\it particular} radial and vertical disk structure assumed here,
the rms variability in the first harmonic is more significant than the rms
variability in the fundamental frequency.  As so little is known, both
observationally and theoretically, about disk vertical profiles we cannot
definitively state whether this feature is generic.  Based upon energy
arguments, however, we {\it can} say that generically disk modes can only
produce rms variability $\sim {\cal O}(1\%)$ if ${\cal H} \sim 1$.  We
regard the question of which frequency, the fundamental or first harmonic,
to associate with the observed 67 Hz feature as an open issue.

Equation (\ref{frequency}) indicates that a 67 Hz (fundamental) modulation
can be produced in a disk around a black hole with a mass of 10.6 $M_\odot$
for a nonrotating hole, and 36.3 $M_\odot$ for a maximally rotating hole.
There are few ``natural" frequencies to invoke in a black hole system.  If
one were to appeal to a process occuring at the marginally stable orbit
($r= 6~GM/c^2$ for a nonrotating black hole), not only would the luminosity
from this region be lower than our mode estimates (due to the no-torque
boundary condition used in disk models), but also the required hole mass
would increase by a factor of at least $16/(3\sqrt{3}) \sim 3$ (if the 67
Hz feature represents the fundamental mode frequency).  Our nonrotating
hole, fundamental frequency, estimate of the required mass would yield an
observed X-ray luminosity (Morgan et al. 1997) $\aproxgt 30\%$ of the
Eddington luminosity, consistent with our assumptions above.

\section{Mode Excitation and Selection Effects}

\subsection{Turbulent Damping and Excitation}

It is possible to use a parametrized stress tensor to estimate the effects
of turbulent viscosity on the modes (NW92, NW93). The canonical energy of a
radial mode is $E_c\sim \sigma^2\rho(\xi_z^2+\xi_r^2)\Delta V$, where
$\Delta V$ is the volume occupied by the mode. Isotropic turbulence
produces a rate of change $dE_c/dt \equiv -E_c/\tau$, with $\tau \sim
|~\alpha \sigma ~[h^2/\lambda_r^2+h^2/\lambda_z^2]~|^{-1}$ and $\lambda_r$,
$\lambda_z$, respectively, being the radial and vertical mode wavelengths.
The corresponding quality factor is given by
\be
{Q_{jn}}^{-1} ~=~ (|\sigma|\tau)^{-1} ~\sim~ \left [j^2+(h/r)n^2 \right ]
     ~\alpha \; ,  
\label{Q}
\ee
as $\lambda_z\sim h/j$ and $\lambda_r\sim \sqrt{hr}/n$, where $j$ and 
$n$ are of order of the number of vertical and radial nodes in any particular 
eigenfunction. Thus, for $\alpha \ll 1$, we can have high mode $Q$.

The other contribution to the fractional width $\Delta f/f$ of the
corresponding feature in a power spectrum comes from the number of
modes that are significantly excited. If we assume that this will
include those whose wavelengths are greater than the maximum eddy size
$L\sim\alpha^{1/2}h$ divided by the corresponding mach number $\sim
\alpha^{1/2}$, equation (\ref{frequency}) gives
\be
\Delta f/f ~\gtrsim~ \Delta n ~(h/r) ~\sim~ \sqrt{h/r} \; . 
\label{Df}
\ee
The minimum effective frequency width will be composed of this span of modes, 
each broadened by $1/Q\sim\alpha$.

The above estimates are for {\it isotropic} viscosity.  If the turbulence
does not efficiently couple to the vertical gradients of the modes, then
the mode $Q$ value is increased by a factor $\sim (j \lambda_r/h)^2$
(NW93).  Aside from damping modes, turbulence can also potentially excite
modes.  Velocity perturbations in the disk, $\delta \vec v$,
are made up of a mode component, $\delta \vec v_M$, and a turbulent
component, $\delta \vec v_T$. Viscous damping arises from
terms of the form $\delta {v_M}_i \delta {v_T}_j$, while
mode excitation arises from terms of the form $\delta {v_T}_i \delta
{v_T}_j$  (NW93).  It is possible to
make simple estimates of the magnitude of the turbulent excitation, and
balance this against the turbulent damping (NW93).  
The modes are excited  to an amplitude of 
$|\xi^z| \sim \alpha (h/{\lambda_r})^{3/2} ~h$, and
$\sim \alpha \sqrt{{\lambda_r}/{h}} ~h$, for isotropic and anisotropic
viscosity, respectively (NW93). If turbulence is playing the
dominant role in damping and exciting the modes, then we have the following
constraints.  For isotropic turbulence, the $Q$ value is only large for
$\alpha \ll 1$; however, this implies a correspondingly small amplitude.
For anisotropic viscosity, not only can we tolerate a larger $\alpha$, 
but we also achieve a larger mode amplitude, although achieving the
observationally required nonlinear mode amplitude is difficult in either case.

\subsection{Negative Radiative Damping}

As a first approximation, we took the modes to be adiabatic.  In reality,
we expect there to be small entropy changes due to various effects, the
most notable one being radiative losses.  If we have a radiation
pressure dominated atmosphere, as is likely in high-luminosity disks, one
properly should use
\be
\Delta P ~=~ \gamma {{P}\over{\rho}} \Delta \rho ~+~ \gamma {{P}\over{s}} 
     \Delta s ~~, 
\label{nonadiabatic}
\ee
where $s$ is the specific entropy.  We can
estimate\footnote{Throughout this section our estimates shall be 
based upon vertically averaged quantities rather than fully $z-$ dependent
quantities.} the effect of  this term for a radiation pressure dominated 
atmosphere, where one has
\be
4 {{P}\over{s}} {{Ds}\over{Dt}} ~\approx~ - {{c}\over{\kappa_{es}}} \vec
      \nabla \cdot {{\vec \nabla P}\over{\rho}} ~\approx~
      {{c}\over{\kappa_{es}}}  \Omega^2 
\label{DeltaS}
\ee
($c$ is the speed of light, $\kappa_{es}$ is the electron 
scattering opacity, and we have invoked vertical hydrostatic equilibrium.)

For perturbations of scalars, the Lagrangian perturbation operator $\Delta$
commutes with total time derivatives $D/Dt$ (cf. Lynden-Bell \& Ostriker
1967). Using this fact, and the fact that for our modes 
$D\Delta s/Dt = i \omega \Delta s$, we can show that
\bea
{{P \Delta s}\over{s}} &=& \left ( 1 - i 
     {{ 4 \kappa_{es} P \omega} \over {c
     \Omega^2 }}  \right )^{-1}  \left ( \Delta P - {{P \xi_r}}
     {{\partial \ln \Omega^2} \over {\partial r}}  \right ) \quad ~
     \nonumber \\
     &=& \left ( 1 - i~  {{4 \omega \tau_{es} h}\over{c
     }}  \right )^{-1}  \left ( \Delta P - {{P \xi_r}}
     {{\partial \ln \Omega^2}\over{\partial r}}  \right ) , \quad ~
\label{PDeltaS}
\eea
where $\tau_{es}$ is the electron scattering optical depth of the
atmosphere.  The term proportional to $\partial \ln \Omega^2/\partial r$ in
the above is negligible in the WKB approximation.  The Lagrangian variation
of the entropy is therefore directly proportional to the
Lagrangian variation of the pressure. 

We do not know the true radial and vertical structure of a disk model that
correctly describes the spectral observations of GRS 1915+105.  However,
for a ``standard" Shakura-Sunyaev $\alpha-$disk model, one can show, with
$\omega \sim \kappa$,  that $\omega \tau_{es} h/c \sim \alpha^{-1}$, and
hence is likely to be $\gg 1$.  Taking this to be the case, we can combine
equation (\ref{nonadiabatic}) and equation (\ref{PDeltaS}) to yield
${{\Delta P}/{P}} \equiv \gamma' {{\Delta \rho}/{\rho}}$, where
\be
\gamma' ~\approx~ \gamma \left (  1 ~+~ i {{\gamma c}\over{ 4 \omega
     \tau_{es} h}} \right ) ~\equiv~ \gamma \left ( 1~+~ i \alpha' \right )
      ~~. 
\label{gammaprime}
\ee
We have subsumed our ignorance of the disk's vertical structure 
into the parameter $\alpha'$ which we expect to be of ${\cal O}(\alpha)$.

To estimate the effects that the non-adiabatic terms have upon our modes,
we can substitute the above into 
equation (\ref{dispersion}) with $\partial^2 W/\partial r^2 \approx - k_r^2 W$
and then expand it about the adiabatic solution.  Specifically, we then
have a dispersion relation that can be written in terms of a function
$G(\omega)=0$, to which we are adding a function $H(\omega)$ due to
radiation damping.  If $\omega_0$ is the unperturbed mode frequency and
$\delta \omega$ is its perturbation, we then have to first order \be
{{\partial G}\over{\partial \omega}} \bigg |_{\omega_0} \delta \omega ~+~
H(\omega_0) ~=~ 0 ~~.
\label{G}
\ee
The mode $Q$ value then becomes
\be
Q ~\equiv~ i {{\omega}\over{\delta \omega}} ~=~ -i {{ \partial G/ 
     \partial \ln \omega |_{\omega_0} }\over H(\omega_0)} 
     ~\sim~ -{{2 \Upsilon}\over{\alpha'}} \left ( {\Omega}\over{c_s k_r}
     \right )^2 ~, 
\label{modeQ}
\ee
where on the right side of the above we have used 
equation (\ref{dispersion}), and have taken the term $\Upsilon
\Omega_\perp$ to be dominant (i.e. $\Upsilon \aproxgt 1$).  
The minus sign indicates mode {\it growth}; that is, the modes are {\it
unstable} to radiative losses. As typically
$k_r^{-1} > h$, $\Upsilon > 1$, and $\alpha < 1$, it is fairly easy to
obtain $|Q| \gg 1$.  

\subsection{Selection Effects}

As discussed in \S2, the modes must approach nonlinearity in order to be
observationally relevant.  If isotropic turbulent excitation determines
the mode amplitude, then there is a natural maximum vertical perturbation
amplitude with ${\cal H} \ll 1$.  If the turbulence is anisotropic, on the
other hand, then it is possible to achieve ${\cal H} \sim 1$, so long as
$\alpha \sim \sqrt{h/\lambda_r} \aproxlt 1$.  For this case we
expect $Q$ values $\sim \alpha^{-3}$.  Radiative excitation
also naturally leads to ${\cal H} \sim 1$.  At some point, currently
unknown, damping effects and nonlinearities must limit this radiative
mode growth. As a zeroth-order approximation we take the linear
growth rates calculated above to be relevant to the near nonlinear regime,
and thus be equal to the saturated damping rate.  The above radiative $Q$
value then gives a rough estimate of the width of observationally relevant
modes.

The low rms variability of the modes means that only modes with high $Q$
values will be detectable.  Observational limits for $Q \sim 20$ were rms
$\sim 0.3\%$, and detectable rms $\propto Q^{-1/2}$ for narrow modes.
Considering the dominant contributions as $n$ and $j$ increase, turbulence
leads to $Q \propto j^{-2}$ (isotropic) and $Q \propto n^{-2}$
(anisotropic), whereas radiative damping leads to $Q \propto j^2/n^2$
The limiting detectable rms variabilty is therefore linear in
$n$ and $j$. Turbulent excitation only leads to high $Q$ values for modes
with low $j$ (isotropic) or low $n$ (anisotropic).  Radiative damping, on
the other hand, only leads to high $Q$ values for modes modes with low $n$
and high $j$.  We note, however, that the mode calculations require a
smooth, well-behaved background to perturb.  If small scale turbulence
truly is responsible for disk viscosity, our unperturbed hydrodynamic
equations are only relevant for the largest scales, again favoring both low
$n$ and low $j$.  For all cases considered above, a high $Q$ value is most
easily achieved for $\alpha \ll 1$.  The 67 Hz mode seen in GRS 1915+105,
if a diskoseismic mode, therefore is likely one with low $n$ and $j$, and
$\alpha \ll 1$.

\section{Conclusions}

The main motivations for attributing the 67 Hz feature in GRS 1915+105 to a
diskoseismic mode are that these modes: 1) are related to a ``natural"
frequency in the disk (i.e. the maximum epicyclic frequency); 2) their
spectra are expected to be characteristic of the inner, hottest regions of
the disk; 3) their frequencies are relatively insensitive to changes in
luminosity; and 4) they have low rms variability.  This latter feature,
although in agreement with the observations, is the strongest constraint.
These modes {\it cannot} be applied to systems that show $\aproxgt 10\%$
rms variability over a wide range of energy bands.  We have identified two
mechanisms, turbulent excitation and negative radiation damping, that
naturally lead to appreciable mode amplitudes with high $Q$ value for
$\alpha \ll 1$.

Again, we stress that the only other ``natural" disk frequency, the
Keplerian rotation frequency at the last marginally stable circular orbit,
would require a black hole mass a factor $\aproxgt 3$ larger.  Furthermore,
less flux is emitted from that region than from the $g-$mode region.

We saw that there were two requirements for the $g-$modes to be
observationally detectable. First, the $g-$modes had to be wide, or
equivalently the disk had to be luminous, assuming that $\sqrt{c_s} \propto
\sqrt{L}$.  Second, the $g-$mode amplitudes had to approach the nonlinear
regime.  Ideally, one should perform numerical simulations of the nonlinear
development of the $g-$modes to determine mode widths, $Q$, and luminosity
modulation more accurately.  It is interesting to note that the MHD
simulations of Stone et al. (1996) do show copious $p-$mode production
associated with the turbulence (Gammie 1996). (MHD simulations have yet to
be performed for rotation and epicyclic frequency profiles relevant to the
$g-$mode trapping region.)

These $g-$mode oscillations should also exist in accretion disks around
compact (soft equation of state), weakly magnetized ($B < 10^8$ gauss)
neutron stars.  Under these conditions, the inner radius of the disk will
be less than those radii where the $g-$mode exists.  Again, large amplitude
(rms $\gg$ 1\%) features {\it cannot} be explained with these modes.

Even if the 67 Hz feature seen in GRS 1915+105 does not turn out to be a
diskoseismic mode, it points out two important lessons.  First, BHC systems
can produce relatively stable, high-frequency features.  Second, the {\it
Rossi X-ray Timing Explorer} is capable of detecting and characterizing
these features despite their weak variability.  The search for diskoseismic
modes in this system and other BHC has therefore become a viable and
worthwhile pursuit.

\acknowledgements

The authors would like to acknowledge useful conversations with Ed Morgan.
M.A.N. was supported by NASA grant NAG 5-3225, R.V.W. by NASA grant NAG
5-3102, D.E.L. by  a National Defense Science and Engineering Fellowship, and
M.C.B. by NSF grants AST-95-29175 and AST-91-20599.

\end{document}